\documentclass[conference]{IEEEtran}
\usepackage{enumitem}
\usepackage{flushend}
\usepackage{graphicx}
\usepackage{subcaption}
\usepackage{soul}

\graphicspath{ {figures/} }

\usepackage[colorinlistoftodos]{todonotes}

\newcommand{\query}[1]{\textit{`#1'}}

\newcommand{\bing}[1]{Bing, a major commercial web search engine}

\def\BibTeX{{\rm B\kern-.05em{\sc i\kern-.025em b}\kern-.08emT\kern-.1667em\lower.7ex\hbox{E}\kern-.125emX}}

\begin{document}

%
\title{Analyzing Web Search Behavior for Software Engineering Tasks}

%

%

\author{
\IEEEauthorblockN{Nikitha Rao, Chetan Bansal, Thomas Zimmermann, \\ Ahmed Hassan Awadallah, Nachiappan Nagappan}
\IEEEauthorblockA{\textit{Microsoft Research} \\
\{t-nirao, chetanb, tzimmer, hassnam, nachin\}@microsoft.com}
}

%
\maketitle
\begin{abstract}
Web search plays an integral role in software engineering (SE) to help with various tasks such as finding documentation, debugging, installation, etc. In this work, we present the first large-scale analysis of web search behavior for SE tasks using the search query logs from Bing, a commercial web search engine. First, we use distant supervision techniques to build a machine learning classifier to extract the SE search queries with an F1 score of 93\%. We then perform an analysis on one million search sessions to understand how software engineering related queries and sessions differ from other queries and sessions. Subsequently, we propose a taxonomy of intents to identify the various contexts in which web search is used in software engineering. Lastly, we analyze millions of SE queries to understand the distribution, search metrics and trends across these SE search intents. Our analysis shows that SE related queries form a significant portion of the overall web search traffic. Additionally, we found that there are six major intent categories for which web search is used in software engineering. The techniques and insights can not only help improve existing tools but can also inspire the development of new tools that aid in finding information for SE related tasks.
\end{abstract}

\section{Introduction}

Internet has become an invaluable source of information in our daily lives. Web search plays a key role in finding relevant information from the internet. A typical knowledge worker frequently searches for data, experts and tools to help with their daily jobs~\cite{reinhardt2011knowledge}. In the software engineering domain, web search is heavily used by developers to assist with various tasks such as finding API documentation~\cite{nasehi2012makes}, code samples~\cite{brandt2010example,stylos2006mica}, installation procedures, solutions to fix bugs and other issues, navigating to resources, etc. 

Several aspects of web search usage and user behavior have been extensively studied in prior work. However, there is limited understanding of web search usage for software engineering (SE). Prior work has majorly focused on code search in particular ~\cite{Bajracharya2012, Stolee:2014:SSS:2628068.2581377, Sadowski:2015:DSC:2786805.2786855} with several tools being built to facilitate code search~\cite{bajracharya2006sourcerer, martie2017understanding, kim2018f,holmes2009developers}. Still, software engineers use web search for several other tasks, besides code search, such as finding tutorials, bug fixes, tools, documentation, discussions, etc. Xia et al.~\cite{Xia:2017} categorized the various tasks performed by software engineers that make use of web search and assessed their frequency and difficulty level by analysing the search logs collected from $60$ developers and additionally interviewed $12$ software engineers. Rahman et al.~\cite{Rahman:2018:EDU:3196398.3196425} analyzed search logs from $310$ Google developers to compare code related and non-code related queries and observed that code related search often required more effort than non-code search.

Understanding usage of web search for software engineering tasks is important not just for improving web search but also to inspire the development of new tools and techniques for software engineering tasks. As discussed in Section~\ref{sec:discussion}, we find that software engineering queries are not as effective as other queries in general web search engines, highlighting the need for custom search engines to enable better search experience for software engineers. The insights generated can also be used to improve existing software products based on the various problems users face. Additionally, the insights from this study can be leveraged to enable personalized search experiences, to improve software tool recommendations and help the users to be more productive.

In this paper, we present the results from the first large-scale study of web search behavior for software engineering tasks using millions of search queries and search sessions from Bing, a commercial web search engine\footnote{The datasets cannot be made public due to GDPR and other privacy laws.}. Firstly, in order to identify SE search queries from other non-SE queries, we use distant supervision to build a machine learning based classifier for distinguishing between SE and non-SE queries. We follow this with a comparative study to better understand the differences in web search behavior for SE and non-SE queries. Subsequently, we propose a taxonomy of intents to identify the various tasks for which web search is being used for software engineering. Following that, we build a heuristics based model to automatically categorize the search intent for a given SE query. Next, we analyze the distribution of intents across various web search metrics such as popularity, success rate and effort estimation. We further look at the trends in intents across sessions, different device types and in the temporal space. Our work is different from prior work as we perform the first large scale study that analyses web search behavior for all software engineering tasks and not just code search. Additionally, our study focuses on the general population as opposed to developers from a single company. Specifically, we make the following contributions: 
\newline
\begin{enumerate}
  \item We propose a distant supervision based machine learning classifier to distinguish software engineering related search queries from other queries (Section~\ref{sec:se-classifier}).
  \item We propose a taxonomy of intents to characterize the web search behavior for software engineering tasks. This taxonomy includes the following intents: API, Debug, HowTo, Learn, Installation, Navigational and Miscellaneous (Section~\ref{sec:taxonomy}).
  \item We perform a comparative study to understand how software engineering related search queries and sessions are different from other queries and sessions by analyzing one million search sessions using the query logs from Bing web search engine (Section~\ref{sec:se-analysis}).
  \item To better understand the distribution of intents in SE queries, we propose a heuristics based intent classification model (Section~\ref{sec:intent-classifier}).
  \item Lastly, we investigate the distribution of intents for six million queries across various web search metrics such as popularity, success rate and effort estimation. We also compare query trends across sessions, device types and in the temporal space (Section~\ref{sec:intent-analysis}).
\end{enumerate} 

The rest of the paper is organized as follows. We start by presenting the related work in Section~\ref{sec:related}. In Section~\ref{sec:data}, we describe the query logs dataset used for our analysis. We follow this with SE query classification in Section~\ref{sec:se-classifier} and discuss the SE query analysis in Section~\ref{sec:se-analysis}. We then describe the taxonomy of intents for SE queries in Section~\ref{sec:taxonomy}. The intent classification and analysis is presented in Sections~\ref{sec:intent-classifier} and~\ref{sec:intent-analysis} respectively. Finally, we present the discussion in Section~\ref{sec:discussion} and conclude in Section~\ref{sec:conclusion}.

\section{Related work}
\label{sec:related}
There has been a significant amount of work from the data mining and information retrieval communities around characterizing and improving web search. In the empirical software engineering community, the primary focus has been on analyzing code search by developers. In this work, we leverage insights and metrics from prior work to better understand web search in context of software engineering. 

\subsection{Web search behavior}
Prior work has extensively studied various aspects of web search behavior. Ong et al.~\cite{oj17-sigir} examined user behavior for Mobile search and Desktop search queries; highlighting several differences in usage including the type of queries and the interaction with the results. Other work focused on characteristics of the results such as the effect of snippet length and content~\cite{mam17-sigir} and the effect of the number of documents in the result list~\cite{ka15-sigir}. There also has been studies that focused on query characteristics such as the query interface and query difficulty ~\cite{akb13-sigir}. These studies are rather generic in nature and aim to provide a general characterization of how people use and interact with web search.

There has also been several studies on web search usage for specific user groups in well defined segments based on domain independent factors such as demographics, task type or task difficulty. For example, Mehrotra et al.~\cite{Mehrotra2017} studied search engine usage across different ages, genders and other demographics. Moreover, the level of difficulty of the task was also shown to have a notable impact on interactions with search engines~\cite{aula2010does,kim2014modeling}. 
Additionally, web search in healthcare and medical diagnosis has also received significant attention~\cite{spink2004study,Schoenherr2014}. Spink et al.~\cite{spink2004study} provided general characterization of medical and health queries. Search query analysis has been also used to understand and characterize user behavior in various domains such as web security \cite{bansal2020ransomware} and e-commerce \cite{rao2020product}.

In this work, we build upon the prior work for analyzing and characterizing web search usage for software engineering tasks. We also discuss the implications of this characterization to improve existing tools and inspire the development of new tools to better support software engineering tasks.

\subsection{Code search}
In the software engineering community, there has been substantial work done in understanding and improving code search \cite{bajracharya2006sourcerer}, \cite{martie2017understanding}, \cite{kim2018f}, \cite{holmes2009developers}. Bajracharya et al. \cite{Bajracharya2012} analyzed usage of Koders.com, a specialized code search engine, by developers. They perform lexical analysis of the search queries and use topic modelling techniques to extract $50$ topics from the search queries. Similarly, Stolee et al. \cite{Stolee:2014:SSS:2628068.2581377} surveyed developers on the tools used for code search and found that $69\%$ of the participants used web search for code search and were not satisfied with the existing code search tools. Sadowski et al. \cite{Sadowski:2015:DSC:2786805.2786855} surveyed $27$ developers at Google to better understand the various properties of code search queries. 

Xia et al.~\cite{Xia:2017} collected search logs from $60$ developers and interviewed $12$ software engineers to categorize search tasks into $34$ buckets across seven different categories. They also carried out a survey to understand the level of difficulty and frequency of these search tasks. Additionally, they found that developers are more likely to search for code on web search engines rather than specialized code search engines. 
Rahman et al.~\cite{Rahman:2018:EDU:3196398.3196425} analyzed search logs from $310$ developers at Google ($150,000$ search queries) and built a statistical model based on Stack Overflow tags to classify search queries into code-related and non-code-related queries. They then compared the search sessions with respect to duration, query length, result clicks, and query reformulations and found that code related searches sessions often requires more effort than general non-code search sessions. Hassan et al.~\cite{hassan2020empirical} mined search logs to characterize the usage of web search for debugging errors and exceptions.

Our work differs from the existing work in the software engineering community in several aspects. This is the first large-scale study, conducted on millions of search queries and sessions, to analyze web search behaviour for all software engineering tasks and not just code search or debugging. Web search is often used for several other software engineering tasks like debugging, navigation, learning and installation. Further, by using the search logs from Bing web search engine, we are able to analyze the behavior of the general population without limiting ourselves to developers in a commercial setting.

\section{Web search logs}
\label{sec:data}

In this study, we use the search query logs from Bing, a commercial web search engine. The logs comprise of a rich set of metadata associated with each user query. Please note that the logs are anonymized to remove any user identifiable information before any analysis was conducted and all results presented in this paper are aggregated over several user queries and sessions.

\subsection{Terminology}
In accordance with the terminology defined in Web search, here is a list of key terms along with their definitions that are frequently used throughout the paper.

\begin{itemize}
    \item \textbf{Search query}: The raw query text that a user enters into the search engine. 
    \item \textbf{Client}: A user facing application used for browsing the search engine and doing search queries. Clients are uniquely identified by using various tracking mechanisms, for instance, browser cookies.
    \item \textbf{Search session}: The various search queries that a user may enter consecutively until there is either a thirty minutes period of inactivity \cite{radlinski2005query} or the browser is closed.
    \item \textbf{Result URLs}: Ordered list of URLs displayed by the search engine in response to the user's search query.
    \item \textbf{Click URLs}: List of URLs which were clicked on by the user from the result URLs ordered by the time of click.
    \item \textbf{{Dwell Time}} - The amount of time spent by the user on the page resulting from a click. Previous research has shown that dwell time has a significant correlation with users satisfaction from the resulting web page~\cite{Fox2005}.
    \item \textbf{{SAT (Satisfactory) Click}} - A click is said to be satisfactory if the user has a dwell time of over thirty seconds as proposed by Fox et al.~\cite{Fox2005}. 
\end{itemize}

\subsection{Scope of the study}
\label{sec:scope}
Web search patterns tend to vary significantly based on several factors such as geography, locales, etc. To minimize this variance, we have limited the scope of this work to en-US locale. Further, we filter out traffic generated by bots since we aim to analyze user behavior in this study. 
\section{SE Query Classification}
\label{sec:se-classifier}

In order to understand how web search is used for software engineering tasks, we first need to be able to distinguish SE search queries from other queries. 

\subsection{Inferring labels}
\label{sec:se-labels}

\begin{table}
\caption{Domains used for inferring labels}
\vspace{-3mm}
\label{table:1}
\begin{center}
 \begin{tabular}{|l|p{4.5cm}|}
    \hline
    \textbf{SE domains} & \textbf{Description} \\
    \hline
    github.com & Largest collaborative software development platform \\ 
    \hline
    developer.mozilla.org & Documentation for web standards and Mozilla projects \\
    \hline
    docs.oracle.com & Documentation for Oracle products including Java \\
    \hline
    developer.android.com & Official website for Android applications and OS \\
    \hline
    stackoverflow.com & One of the most popular question and answer website for programmers \\
    \hline
 \end{tabular}
\vspace{-6mm}
\end{center}
\end{table}

A significant amount of labeled data containing both SE queries and non-SE queries is required to build a high accuracy machine learning based classifier. However, getting the labels can be quite laborious and expensive. Hence, we propose the use of distant supervision techniques to build our training dataset. We use the website listings from Alexa.com \cite{Alexa} and the click information from the query logs and perform the following steps:
\begin{enumerate}
    \item We collect the top $5$ SE related websites from the software category on Alexa.com (summarised in Table ~\ref{table:1}).
    \item We then process $5$ days of the query logs described in Section~\ref{sec:data} and sample $1$ million queries. We discuss the data sampling process in detail in Section~\ref{sec:se-train-data}.
    \item Subsequently, we label the subset of sampled queries having at least one click on a SE domain (listed in Table \ref{table:1}) as SE related. The remaining randomly sampled queries are labeled as Non-SE related.
\end{enumerate}
The key insight leveraged here is that the queries which result in a click on the SE related domains are generic and diverse enough to train a more generic classifier for all SE related queries. The efficacy of this method is confirmed by the results of the evaluation process. An additional optimization was performed to remove navigational queries from the data by removing queries which resulted in a click to the home page of the SE domains. Lastly, a machine learning model was used to train the classifier rather than solely relying on the heuristics, as (i) Not all queries with clicks leading to these websites may be SE related, for example, the queries for logging in and financial statements, (ii) The top $5$ SE related websites are a small subset of the SE related websites.

\begin{figure*}
\caption{Top 20 positively and negatively correlated features}
\vspace{-3mm}
\includegraphics[width=\textwidth]{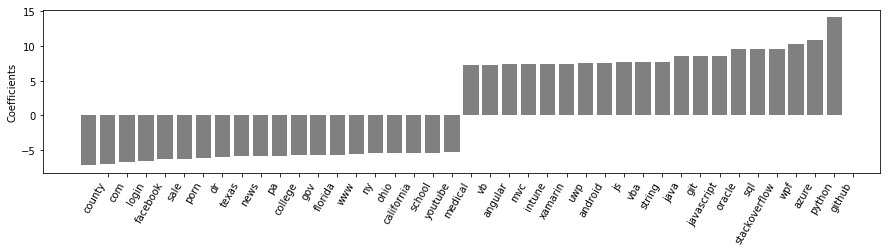}
\vspace{-10mm}
\label{figure:1}
\end{figure*}

\subsection{Training data}
\label{sec:se-train-data}
As described earlier, there is a rich set of metadata associated with the search queries such as the query text, result URLs, clicked URLs, dwell times, etc. However, we only use the features derived from the query text for training the classifier to ensure accurate classification despite the absence of other information. For instance, not all queries lead to clicks on the search results. Also, since our heuristics rely on the clicked URLs, they are explicitly removed from the model features to prevent overfitting. The query text is preprocessed to remove all non-alphanumeric characters. This followed by a transformation of the query text to vector form by using the TF-IDF representation. The TF-IDF representation ensures that the model does not overfit on frequently occurring words such as stop words. 

For training and testing of ML classifiers, around $2$ million search queries are sampled from a time period between May $1^{st}$, $2020$ and May $15^{th}$, $2020$. Presently, the queries are limited to en-US and normal traffic as described in Section~\ref{sec:scope}. The queries are sampled using stratified sampling techniques with a $1:10$ ratio of SE:Non-SE queries since SE queries form a small fraction of the overall web search traffic. Considering that the extract ratio of SE:Non-SE queries in the real world is not known, a low sampling ratio is chosen arbitrarily. Accordingly, the dataset contains $2$ million non-SE and $0.2$ million SE queries. Lastly, a $70:30$ random split of the dataset is used to generate the train and test datasets.

\begin{table}
\begin{center}
\caption{10-fold cross-validation comparison}
\vspace{-1mm}
\label{table:2}
 \begin{tabular}{|c|c|c|c|c|}
 \hline
 \textbf{Model} & \textbf{Precision} & \textbf{Recall} & \textbf{F1} & \textbf{AUC} \\
 \hline
 AdaBoost & 0.891 & 0.516 & 0.653 & 0.826 \\
 \hline
 DecisionTree & 0.911 & 0.890 & 0.900 & 0.935 \\
 \hline 
 \textbf{LinearSVC} & \textbf{0.941} & \textbf{0.920} & \textbf{0.930} & \textbf{0.989} \\
 \hline 
 LogisticRegression  & 0.940 & 0.891 & 0.915 & 0.988 \\
 \hline
\end{tabular}

\end{center}
\vspace{-6mm}
\end{table}

\subsection{Model selection}
The task of identifying SE and non-SE queries is formulated as a binary classification problem. In order to find the most suitable classification model, several experiments are run with different machine learning algorithms, namely, AdaBoost, Decision Trees, Logistic Regression, Linear SVC (Support Vector Classification). The Scikit-learn 0.20.0 package for Python 3.7.1 is used for training and evaluating all the classifiers. Note that our goal is to explore the feasibility of classifying SE related search queries rather than finding the best-fitting classification model. Hence, the default hyper-parameters for these classifiers are used as is.

The classifiers are evaluated and compared against the following four widely used metrics: Precision, Recall, F1-Score and AUC (area under ROC curve). 
In line with existing work such as \cite{kohavi1995study}, the classification models are evaluated using $10$-fold cross-validation. Table \ref{table:2} summarises the performance of classifiers based on the evaluation metrics. We note that the LinearSVC classifier outperforms all other models with a F1-score of $0.93$ and AUC of $0.989$. Consequently, the LinearSVC model is chosen to classify the queries as SE and Non-SE.

\subsection{Model evaluation}
The efficacy of the LinearSVC model for classifying SE and non-SE queries is further evaluated in the following ways. Firstly, an additional analysis of the model on the automatically labeled test dataset described in the previous section is carried out. This is followed by an evaluation of the model on manually annotated data. Lastly, a qualitative analysis of the feature weights learnt by the model shows that the model is highly generic and does not overfit on the hueristics used.

\begin{table}
\caption{Evaluation results of LinearSVC model on test data}
\vspace{-3mm}
\label{table:3}
\begin{center}
 \begin{tabular}{|c|c|c|c|c|} 
 \hline
 \textbf{Class} & \textbf{Precision} & \textbf{Recall} & \textbf{F1-score} & \textbf{Support} \\
 \hline
 SE & 0.94 & 0.93 & 0.93 & 149,558 \\
 \hline
 Non-SE & 0.98 & 0.98 & 0.98 & 450,442 \\  
 \hline
\end{tabular}
\end{center}
\vspace{-6mm}
\end{table}

\textbf{Evaluation on inferred labels} - As shown in Table \ref{table:2}, the LinearSVC model has high accuracy on the test data created using the inferred labels (described in Section~\ref{sec:se-labels}). Given that the data has a class imbalance, the metrics are individually calculated for the two classes, SE and Non-SE, and the resulting scores are reported in Table \ref{table:3}. Here, support refers to the number of samples that belong to a given class. We find that the LinearSVC model has an F1-score of $0.93$ for the SE class (minority) and we achieve upto $0.98$ F1-score for Non-SE class (majority). Hence, validating that the model can classify both SE and Non-SE queries in the test set with high accuracy.

\textbf{Manual evaluation} - So far, the evaluations were performed on the automatically labeled dataset. To ensure that the classifier does not overfit by simply learning to distinguish search queries leading to the SE websites listed in Table \ref{table:1} from queries leading to clicks on other websites, a manual evaluation is conducted on $200$ randomly sampled search queries from the test dataset. Two of the authors manually and independently annotate the data with SE and Non-SE labels. The Cohen Kappa \cite{cohen1960coefficient} coefficient is then used to measure the inter-rate agreement with a resulting score of $0.91$. Upon evaluating the trained ML model on this manually labeled dataset, we see that the model achieves an accuracy of $0.93$ which proves that the model is generalizable, highly accurate and does not overfit on the training data.

\textbf{Feature coefficients} - The ML classifier is trained using the unigram features extracted from the search queries. The trained model is further analyzed by extracting the top features learned. Figure \ref{figure:1} plots the top $20$ features and their corresponding coefficients learned by the model. It is observed that both the positively (for instance: python, github, string) and negatively correlated features (for instance: county, news, porn) were highly generalized topics that corresponds to SE and Non-SE tasks respectively.

\section{SE Query Analysis}
\label{sec:se-analysis}

In this section, a comparative study is carried out to analyse the differences in SE and Non-SE queries across several metrics that are broadly divided into the following categories: query characteristics (such as counts, query similarity and reformulation metrics), interaction characteristics (such as clicks and dwell time) and geographical trends. This provides key insights into differences in web search behaviour for SE specific tasks and other Non-SE tasks. Table \ref{table:session-metrics} and Table \ref{table:query-metrics} summarise the findings for session-level behaviour and query-level behaviour respectively.

\subsection{Data}
\label{sec:se-dataset}
The analysis is carried out on one million search sessions that are randomly sampled from a time period between May $1^{st}$ and May $15^{th}$, $2020$. Necessary filters are applied to remove automated traffic from bots and services and the scope is limited to users having English locale in the US region. The resulting dataset contains $4,103,219$ queries from $985,920$ distinct clients with $2.1\%$ of the total queries automatically labeled as SE queries using the trained classifier. Additionally, a given search session is labeled as SE related if a majority of queries in the session are SE related. This results in $2.61\%$ of all the web search sessions from the sampled data being labeled as SE search sessions. Thereby validating that web search is significantly used for SE related tasks (Table \ref{table:session-metrics}).

\begin{table}
\caption{Comparison of SE and non-SE search sessions}
\label{table:session-metrics}
 \begin{tabular}{|p{3cm}|l|l|}
 \hline
 \textbf{Metric} & \textbf{SE ($\pm$ SEM)} & \textbf{Non-SE ($\pm$ SEM)} \\ 
 \hline
 Unique Session \% & 2.611 & 97.389 \\ 
 \hline
 Unique Client \% & 2.832 & 97.168 \\ 
 \hline
 Avg. unique query count & 2.186 ($\pm$ 0.014) & 3.05 ($\pm$ 0.005) \\ 
 \hline 
 Avg. similar query \% & 4.278 ($\pm$ 0.072) & 3.555 ($\pm$ 0.01) \\
 \hline
 Avg. term addition count & 2.484 ($\pm$ 0.068) & 2.183 ($\pm$ 0.021) \\
 \hline
 Avg. term removal count & 2.184 ($\pm$ 0.085) & 1.879 ($\pm$ 0.025) \\
 \hline
\end{tabular}
\vspace{-2mm}
\end{table}

\begin{table}
\caption{Comparison of SE and non-SE search queries}
\label{table:query-metrics}
 \begin{tabular}{|p{2.5cm}|l|l|}
 \hline
 \textbf{Metric} & \textbf{SE ($\pm$ SEM)} & \textbf{Non-SE ($\pm$ SEM)} \\ 
 \hline
 Avg. word count & 5.245 ($\pm$ 0.017) & 3.807 ($\pm$ 0.002) \\ 
 \hline
 Avg. character count & 30.521 ($\pm$ 0.0111) & 24.088 ($\pm$ 0.0163) \\ 
 \hline
 Avg. click count & 0.41 ($\pm$ 0.002) & 0.449 ($\pm$ 0.003) \\ 
 \hline 
 Avg. SAT click count  & 0.217 ($\pm$ 0.001) & 0.236 ($\pm$ 0.002) \\
 \hline
 Avg. total dwell time \newline (in sec) & 270.051 ($\pm$ 2.072) & 307.549 ($\pm$ 0.339) \\
 \hline
\end{tabular}
\vspace{-4mm}
\end{table}

\subsection{Session-level analysis}

\noindent\textbf{Number of unique queries}: While query characteristics like the number of unique queries provides a good estimate of query popularity, it can also be used as a proxy to estimate the session length. A higher number of unique queries per session indicates that the user is required to explore different variations of the search query before their information need is fulfilled resulting is longer search sessions. Note that using the unique query counts ensures that the same query is not accounted for multiple times even if the user refreshes the search page or hits the back button. It is observed from Table \ref{table:session-metrics} that SE search sessions contain an average of $2.186$ unique queries which is $28.33\%$ lower than non-SE search sessions demonstrating that the SE search sessions tend to be shorter.\\

\noindent\textbf{Query similarity}: Another useful query characteristic in web search is the diversity of queries within a given session. To explore the level of diversity in a given session, the similarity of every subsequent query is measured against the first query. This helps to capture the evolution of query formulation as the session progresses. The query similarity can additionally reflect on whether a user was successful in their search task. An unsuccessful search session might lead to strong similarity between the initial query and the corresponding queries with terms being added or removed as the session progresses. 

Prior to computing the similarity score between a pair of queries in the session, a standard text normalization is carried out where the query text is converted to lowercase, any extra white-space characters and stop words are eliminated. This results in a bag-of-words representation of the query terms. Finally, the similarity score for any two queries is computed using the Jaccard coefficient between the two sets of bag-of-words. It is observed from Table \ref{table:session-metrics} that SE search sessions contain $27.17\%$ more similar search queries than other Non-SE sessions implying that users browse related topics at a higher rate in SE sessions.\\

\noindent\textbf{Reformulation strategies}: While the number of unique queries in a session and the similarity between queries shed light on the length of the sessions and track the evolution of queries as a session progresses. We further analyse the strategies employed by a user when transitioning from one query to another. Reformulated queries refer to a pair of queries having a similarity larger than a set threshold. The reformulation of a query can take place in several ways. However, for the sake of simplicity we look at the following two methods: (i) \textit{Term Addition}, where one or more words are added to the query, and (ii) \textit{Term Removal}, where one or more words are removed from the query. It can be observed from Table \ref{table:session-metrics} that SE search sessions have a significantly higher rate of term additions and removals. This is consistent with our previous finding of SE sessions having a higher percentage of similar search queries.

\subsection{Query-level analysis}

\noindent \textbf{Word and character counts:} The simple query characteristics based on word and character counts are used to better understand the formulation of SE and Non-SE queries. It can be observed from Table \ref{table:query-metrics} that SE queries have a $37.8\%$ higher word count on average when compared to Non-SE queries. Additionally, it can be observed that SE queries contain $27.17\%$ more characters on average. Based on a manual analysis, we observe that a majority of the SE queries concern with debugging tasks with searches having extremely descriptive error messages that lead to high word and character counts.\\

\noindent \textbf{Number of clicks:} The level of interaction in terms of number of result URLs clicked may vary significantly based on the type of query with some tasks requiring more effort in finding the relevant information than others. Using the click information for a given query being logged in the search logs, the average number of clicks is computed for SE and Non-SE queries. To ensure that the clicks are relevant, non-result clicks (such as, clicks on sponsored results) and clicks that lead to another search result page (such as, spelling corrections, related search clicks, etc.) are removed.
It can be observed from Table \ref{table:query-metrics} that SE search queries have a lower click rate as well as a lower SAT click rate than non-SE search queries demonstrating that SE related search tasks are generally more difficult than other search tasks.\\

\noindent\textbf{Dwell time:} The difference in dwell time on clicked results is another interesting interaction characteristic that we explore. Prior work has shown that the amount of time spent by users on the clicked document is an important indicator of whether they are satisfied with the content~\cite{Fox2005}. The dwell times, recorded for each click, are averaged over the given query. It can be observed from Table \ref{table:query-metrics} that the total dwell time on average for SE search queries has a $13.8\%$ reduction compared to other queries thereby demonstrating that SE queries are shown to be not as successful as other Non-SE queries. \\

\begin{figure}
\begin{center}
\caption{Normalized SE queries distribution in US}
\vspace{-3mm}
\includegraphics[width=\linewidth]{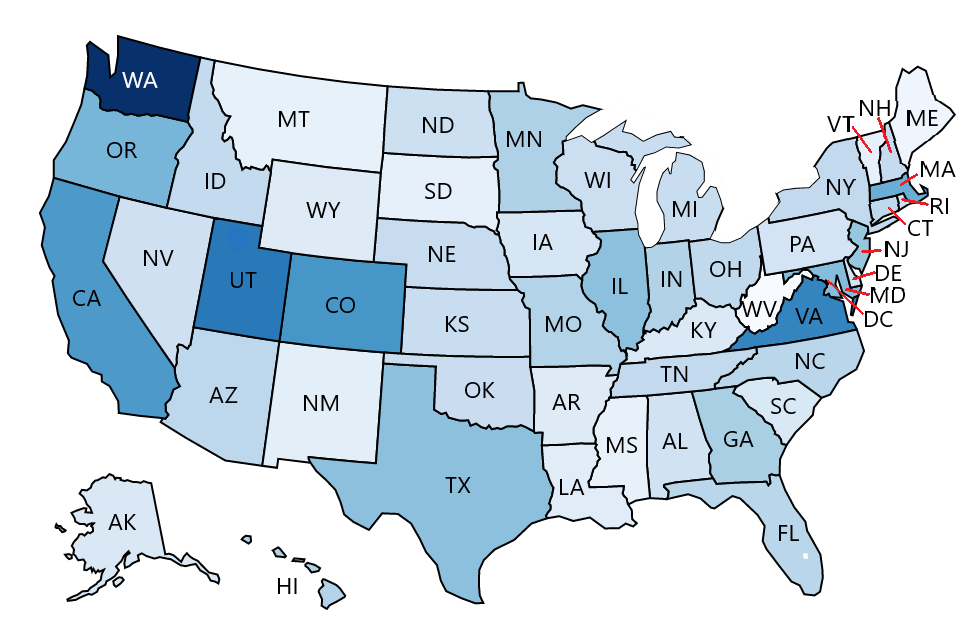}
\vspace{-10mm}
\label{fig:geographical-trends}
\end{center}
\end{figure}

\noindent\textbf{Geographic trends:} To better understand how the search trends vary across different states in the US, a heat map of the normalized SE query volume, computed as the ratio of SE queries to that of Non-SE queries is plotted. The normalization removes any bias that may occur from states that are large (in population or internet penetration), like California, Texas, New York, having high web search activity. As seen in Figure~\ref{fig:geographical-trends}, Washington has the highest SE query volume closely followed by Utah, California, Virginia, Colorado, Oregon and Texas. These results seem to correlate with the statistics provided by the U.S. Bureau of Labor Statistics in the \textit{Occupational Employment and Wages Report} \footnote{https://www.bls.gov/oes/current/oes151256.htm\#st} from May 2019 that measure employment rate, skill and income.


\section{Query taxonomy}
\label{sec:taxonomy}

\begin{table}
\begin{center}
    \caption{Taxonomy of user intent in prior work.}
    \label{tab:prior_taxonomy}
    \begin{tabular}{|p{2.4cm}|p{5.5 cm}|}
        \hline
        \textbf{Reference} & \textbf{Taxonomy} \\
        \hline
        Broder et al.~\cite{Broder} & Informational, Navigational, Transactional \\
        \hline
        Fourney et al.~\cite{Fourneyinproceedings} & Operation Instruction, Troubleshooting, Download, Reference, General Information, Off-Topic \\
        \hline 
        Panichella et al.~\cite{7332474} & Feature Request, Opinion Asking, Problem Discovery, Solution Proposal, Information Seeking, Information Giving \\
        \hline
        This work & API, Debug, HowTo, Installation, Learn, Navigational, Miscellaneous \\
        \hline
    \end{tabular}

\end{center}
\vspace{-6mm}
\end{table}

\begin{table*}

  \begin{center}
    \caption{Intent categories for SE queries.}

    \label{table:taxonomy}
    \begin{tabular}{|p{2cm}|p{7cm}|p{7cm}|}

    \hline
    \textbf{Intent} & \textbf{Description} & \textbf{Examples} \\
    \hline
    API &  Queries where the user wants to learn more about a specific API element in the software. These queries often lead to the documentation page of the API. & `bitflyer rest api', `reshape pandas', `cpp stdvector', `fwrite linux', `keras attention tf20', `flex css', `htm5 drag and drop', `excel text functions', `break matlab', `dotnet core web api docker', `microsoft graph api', `sqlbulkcopy data table' \\
    \hline
    Debug & Queries related to debugging an error or issue which typically include error messages, parts of stack traces, and sometimes a short description to given context to the error. & `dban error disk not found', `gpg error no usable configuration', `run time error arch linux', `cant find mount source devdisk openmediavault', `error code 126 dll cannot be loaded', `steam vr failed initialization', `cant connect to net extender' \\
    \hline
    HowTo & Queries where the user is trying to accomplish a specific task. These queries often contain a short description of what the user wants to accomplish and in some cases the technology they want to use. & `How to copy formula down the column', `selecting rows from dataframe with value at column output row names in r', `how to specify port number in ssms', `ping a server', `use api to pull data from site', `aws amplify add codegen' \\
    \hline
    Learn & Queries where the user is trying to learn about an abstract topic related to software engineering. They can also include queries that are comparing two software or reviewing them. & `keras vs pytorch', `ios opencv', `file system implementation', `access database tutorial', `what is the float property of css', `difference between primary and unique key sql', `c language pointer to function', `radeon vii fps benchmarks' \\
    \hline
    Installation & Queries where the user aims to install a software, tool, package, etc. Often these searches include the target environment, version numbers. & `installing webroot', `ubuntu eigen3 install', `npm install', `microsoft sql server localdb 2016 install', `install r260 driver ', `sql server managment studio download', `visual c++ latest download', `install zoom app for windows 10'  \\
    \hline
    Navigational & Queries where the search engine is used to navigate to a specific resource or web-page. The user has a specific destination in mind and uses the search engine as a shortcut to get there. & `robot mesh studio', `sharepoint', `anaconda', `raspberry pi', `cache camper', `aws.amazon', `url encoder', `typing test', `obs studio', `input mapper', `unity performance', `linux', `eclipse', `azure portal', `switcher studio', `rarbg index page' \\
    \hline
    Miscellaneous & Queries for which none of the above intents were suitable due to insufficient context available to make a decision. & `edge popup allow', `bootstrap prev next tabs', `laps gpo files', `webknight application firewall alert', `free proxy server', `xml dat', `openshift client certificate', `web html'\\
    \hline

    \end{tabular}
  \end{center}
    \vspace{-6mm}
\end{table*}

Software developers can have various intents when searching the web for SE queries. They may want to learn more about a technology, debug an error message they encountered, install a new software and so on. In this section, we aim to understand the various intents associated with SE queries.

To this end, $200$ SE queries are uniformly sampled based on the query length (the number of tokens present in the query) from the SE query dataset generated in Section~\ref{sec:se-dataset}. Three annotators independently inspected all the queries along with the click URLs. Using the open coding approach, they first assigned a label based on what they thought the most prominent intent behind the query was. This was then followed by a discussion to understand the various intent categories and they settled on the following intent categories: \textit{API, Debug, HowTo, Installation, Learn, Navigational} and \textit{Miscellaneous}.

The three annotators then labeled another set of $200$ SE queries, uniformly sampled based on query length, in order to validate the intent categories and to make sure no new categories emerged. The inter-annotator agreement was then computed using Fleiss kappa ~\cite{fleiss1971measuring} with a resulting score of $0.71$ indicating substantial agreement among the annotators. When the raters disagreed, it was either because of lack of context (for example, \query{laps gpo files}, \query{xml dat}) or for queries where multiple categories were applicable (for example, \query{how to update android version}, \query{opencv ios}). A detailed description of the resulting taxonomy along with example queries can be found in Table~\ref{table:taxonomy}.

Table~\ref{tab:prior_taxonomy} summarizes the user intent taxonomies developed in prior works including, general web search~\cite{Broder}, search in the context of interactive applications~\cite{Fourneyinproceedings} and using feedback in the form of review comments in apps~\cite{7332474}. The similarities and differences between our taxonomy and the taxonomies defined in prior works are elaborated below:  
\begin{itemize}
    \item The Navigational intent that Broder et al~\cite{Broder} introduced for general web search is retained and Transactional is rebranded to Installation as it is more suited in the software engineering context. The Informational intent is dissected into finer intent categories that are specific to software engineering like API, Debug, HowTo and Learn. 
    \item A significant overlap is observed in the taxonomy defined by Fourney et al.~\cite{Fourneyinproceedings} for interactive applications. A one-to-one mapping is identified between the some of the intents as Operation Instruction, Troubleshooting, Reference, Download and General Information directly map to HowTo, Debug, API, Installation and Learn respectively. This indicates that the web search intents observed in the context of interactive applications forms a subset of web search intents we observe in software engineering.
    \item Panichella et al.~\cite{7332474} defined a taxonomy for user feedback in the form of reviews for apps. While there was some overlap between intents with Problem Discovery mapping to Debug and Opinion Asking and Information Seeking mapping to Learn or API, it is observed that intents like Feature Request, Information Giving and Solution Proposal are not found in the context of web search for software engineering. 
\end{itemize}

\section{Intent Classification}
\label{sec:intent-classifier}

One of the goals of this study is to analyse the distribution of intents in SE queries. Given the large number of SE queries, we propose a heuristics based classification model to automatically identify the search intent given the search query. The $200$ manually labeled queries (train samples) along with their click URLs from Section~\ref{sec:taxonomy} are analyzed to identify various patterns correlating to each intent class. These include frequently occurring keywords in both the query string and the clicked URLs, number of URLs clicked, type of URLs, etc.

A rule based model, that leverages the heuristics, is then built to automatically infer the intent labels. The performance of the model is then tested on the train samples. The heuristics are iteratively updated until an accuracy of over 90\% is attained on the train samples. The final set of heuristics associated with each intent is described in Table~\ref{table:intent-heuristics}. The precedence order followed by the model, in the decreasing order of specificity, is as follows: Debug, Installation, Learn, HowTo, API, Navigational, Miscellaneous.

To ensure that the model does not overfit on the training samples that was used to tune the classifier, the model is further evaluated on test set of $200$ randomly sampled queries. Upon doing a manual evaluation, it is noted that the model achieves a test accuracy of $93.5$\%. The misclassifications were either due to lack of sufficient context in the query string (for example, \query{CrtDbgBreak return true}, \query{edge popup allow}) or due to multiple intents being applicable for a given query (for example, \query{dns server settings}, \query{how to update anaconda}).

\begin{table}
\small
\vspace{-2mm}
  \begin{center}
    \caption{Heuristics associated with each intent.}
    \label{table:intent-heuristics}
    \begin{tabular}{|p{2cm}|p{5.5 cm}|}
    \hline
    \textbf{Intent} & \textbf{Heuristics} \\ \hline
    API & Keywords - `api', `function', `method', `call', `reference', `ref', `doc', `command' \\ 
    \hline
    Debug & Keywords - `error', `troubleshoot', `fail', `debug', `exception', `care', `fix', `problem', `diagnose', `not working', `solve', `not', `couldnt', `wouldnt', `wont', `cant' \\ 
    \hline
    How-to & Keywords - `how', `question' \\ 
    \hline
    Installation & Keywords - `download', `install', `purchase', `buy'  \\ 
    \hline
    Learn & Keywords - `tutorial', `wiki', `learn', `why', `what', `where', `does', `review', `vs', `versus', `difference' \\ 
    \hline
    Navigational & Single URL click leading to the target website.\\ 
    \hline
    Miscellaneous & None of the above heuristics are satisfied \\ 
    \hline
    \end{tabular}
  \end{center}
  \vspace{-7mm}
\end{table}

\section{Intent Analysis}
\label{sec:intent-analysis}

To better understand the distribution of intents in SE queries, the best performing SE query classifier (i.e. LinearSVC) from Section~\ref{sec:se-classifier} is used to identify the SE queries from a two week time span between May $1^{st}$, $2020$ and May $15^{th}$, $2020$. $6$ million SE queries are then sampled along with other query attributes like click URLs, click counts, request time, dwell times etc for understanding users' behavioural patterns. After which, the heuristics based intent classification model (described in Section~\ref{sec:intent-classifier}) is used to infer the intent labels. This is followed by an analysis on distribution of intents across various metrics like popularity, success rate and effort estimation. Additionally, an analysis on various trends across sessions, different device types and in the temporal space is carried out. The insights gained from this analysis can be leveraged to better understand the web search behaviour of users for SE tasks allowing others to find answers to questions like `what are some frequent issues users are facing with a given software?', `how does the given software tool compare with it's competitor tools based on whether they are have higher success rates and lower estimated effort?', `finding the popularity of different software technologies based on users' web search engagement' and so on. \\

\noindent\textbf{Popularity:} For a given intent, the intent popularity is defined as the percentage of SE queries having that specific intent. It can be observed from Table \ref{tab:intent-metrics} that the most popular intent is Debug closely followed by HowTo, Learn and API; whereas Installation and Navigational queries are far less popular.\\

\noindent\textbf{Success rate:} Fox et al.~\cite{Fox2005} proposed that a search is successful if the user has a dwell time of over $30$ seconds on the clicked URL page. We build on this definition and consider a search to be successful if the dwell time on the last clicked URL is more than $30$ seconds owing to the fact that the user may click on multiple URLs during the search. This is based on the assumption that the user stops looking at other websites once they have found the information they desired. 

For a given intent, the success rate is defined as the percentage ratio of number of successful queries to the total number of queries belonging to that intent category. It can be observed from Table \ref{tab:intent-metrics} that HowTo queries have the highest success rate of $51.33\%$, closely followed by Learn and API; whereas Navigational queries are the least successful. The low success rate for Navigational queries is a result of users spending less than $30$ seconds after navigating to the URL.\\

\begin{table}
\begin{center}
\vspace{-2mm}
    \caption{Comparison of intent metrics.}
    \label{tab:intent-metrics}
    \begin{tabular}{|l|c|c|c|}
        \hline
        \textbf{Intent} &  \textbf{Popularity} & \textbf{Success Rate}  & \textbf{Estimated Effort}\\
        \textbf{} & \textbf{(\%)}  & 
        \textbf{(\%)} & \textbf{(Relative Scale)}\\
        \hline
        API & 17.74 & 49.68 & 33.99x \\
        \hline
        Debug & 21.53 & 48.95 & 11.33x \\ 
        \hline
        HowTo & 19.01 & 51.33 & 21.49x \\ 
        \hline
        Learn & 18.46 & 50.19 & 10.05x \\
        \hline
        Installation & 11.99 & 46.11 & x \\ 
        \hline
        Navigational & 11.27 & 41.58 & 15.82x \\ 
        \hline    
    \end{tabular}
\end{center}
\vspace{-4mm}
\end{table}

\noindent\textbf{Effort estimation:} The effort required to complete a search is estimated to be proportional to the average dwell time for all clicked URLs ~\cite{effort-estimation}. For a given intent, the estimated effort is computed as the mean average dwell time of all queries belonging to that intent category. This score is transformed to a relative scale by representing each score as a factor of the estimated effort for the Installation intent category. 

It can be observed from Table~\ref{tab:intent-metrics} that Installation has the least estimated effort whereas intents like HowTo and API require significantly more effort. This results from the fact that queries with the former intent are generally specific (for example, \query{download anaconda for windows 10}, \query{install zoom app}) whereas the later intents have a rather elaborate course of action with the user carrying out the steps required to accomplish a task (for example, \query{how to filter rows based on column values pandas dataframe}, \query{how can we ping a server}) or trying to understand the documentation of an API (for example, \query{dotnet core web api docker}, \query{sqlbulkcopy data table}). \\

\begin{figure}
\begin{center}
\vspace{-2mm}
\caption{Co-occurrence of intents within sessions.}
\includegraphics[width=\linewidth]{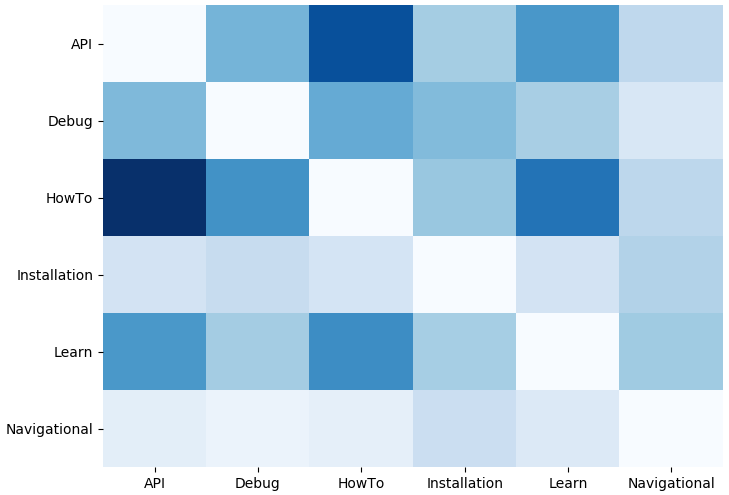}
\label{fig:cooccurrence}
\end{center}
\vspace{-12mm}
\end{figure}

\noindent\textbf{Co-occurrence of intents within sessions:} To better understand the behavioural patterns of users searching for SE queries, a session wise analysis on the temporal co-occurrence of different intents is performed and the findings are aggregated across all sessions. From Figure~\ref{fig:cooccurrence} it can be observed that while less than $1\%$ of Installation and Navigational queries are followed by other intents, HowTo queries are commonly followed by API ($33\%$), Learn ($24.5\%$) and Debug ($20.6\%$) indicating that once the user has figured out how to complete the task at hand they are either trying to learn more about some of the steps involved or they are trying to resolve an issue they encountered when trying to execute the steps. Similar co-occurrences are observed with API queries being frequently followed by HowTo ($29\%$), Learn ($20\%$) and Debug ($15.7\%$). However, it is interesting to note that while Learn is frequently followed by API ($19.8\%$) and HowTo ($21.3\%$), it is rarely followed by Debug ($8\%$). Another interesting insight we observed is that Installation queries are often preceded by Debug ($14\%$) indicating that the user's errors were resolved by installing missing packages. Additionally, it can be seen that Installation is preceded by HowTo ($12.9\%$), API ($11.7\%$) and Learn ($11.5\%$) indicating that users are truly interested in engaging with the software and hence install it. \\


\begin{figure}
\begin{center}
\caption{Hourly distribution of intents.}
\includegraphics[width=\linewidth]{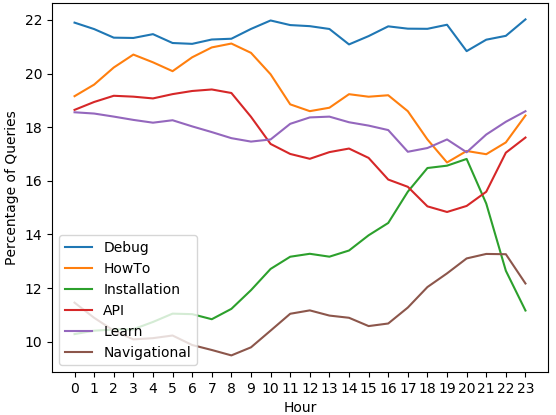}
\label{fig:hourly-trends}
\end{center}
\vspace{-12mm}
\end{figure}

\noindent\textbf{Hourly trends:} To study the variations in temporal trends across different intents, the distribution of intents at every given hour of the day is plotted as the ratio of queries for a given intent to total SE queries at the given hour. From Figure~\ref{fig:hourly-trends} it can be observed that while the distribution of intents like Debug and Learn are roughly the same throughout the day, the other intents distributions show interesting patterns. HowTo and API queries show a significant dip in volume during night time whereas Installation and Navigational queries steadily increase throughout the day and peak at night time. This could mean that users tend to perform more intellectually challenging tasks like learning about APIs or trying to figure out how to use some feature in a software during the day and reserve the less challenging yet potentially time consuming tasks such as downloading and installing software tools or packages for the night time. \\

\begin{figure}
\begin{center}
\caption{Distribution of intents across devices.}
\includegraphics[width=\linewidth]{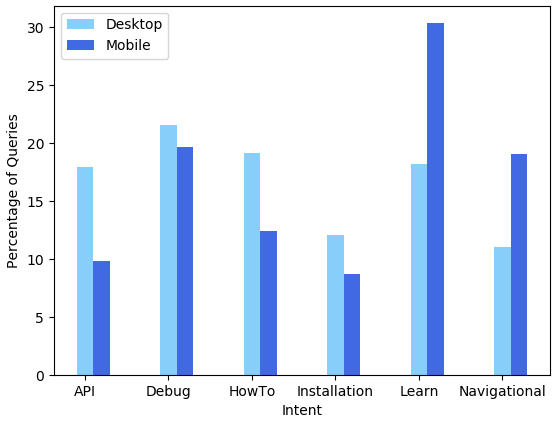}
\label{fig:device-trends}
\end{center}
\vspace{-10mm}
\end{figure}

\noindent\textbf{Device type:} Web search is heavily used by a wide variety of users across different devices like desktop, mobile, etc with varying search usage patterns. Kamwar et al. \cite{kamvar2006large} showed that web search patterns and usage differ vastly between different client form factors like desktop, mobile, etc. To better understand the distribution of SE query intents across desktop and mobile users, the percentage of queries belonging to each intent for the given device is plotted. Looking at Figure~\ref{fig:device-trends}, it is interesting to note that intents like Debug, HowTo, API and Installation are far more popular in desktop clients whereas Learn and Navigational are significantly more popular among mobile clients.

\section{Discussion}
\label{sec:discussion}

There are several implications of this work. First and most notably, the analysis of $1$ million web search sessions in Section~\ref{sec:se-analysis} suggests that \textbf{software engineering related queries are less effective than other types of queries} as a result of higher rates of query reformulations, fewer clicks, and shorter dwell time. Custom search engines may enable better search experience for software engineers. 
The query taxonomy presented in Section~\ref{sec:taxonomy} serves as guidelines for the various modes that need to be supported by search engines to improve the search experience for software engineers.

The search data provides a \textbf{pulse of what software engineers are searching for and what problems they face} on a large scale. This data can be analyzed to generate insights to help improve software products. For example, it allows us to identify the frequent problems users face with a software technology, it can be used to compare different software technologies based on search query properties, or can be used to predict the satisfaction of developers with specific software technology. All of which act as avenues for future work. This information could also be looped back to the creators and users of software technologies, similar to tools like Google Trends.

Lastly, the search history of individuals can be used to \textbf{provide context} to personal assistants such as Siri, Cortana, etc. as well as software bots \cite{softwarebots} to enable personalized search experiences and better software tool recommendations. Integrating context and task-aware search into software tools can be used to improve the productivity of users. Additionally, search data can also be used as signal for detecting the task type~\cite{Bao2018}. 

\section{Conclusion}
\label{sec:conclusion}
In this paper, we presented the first large scale study on web search usage for software engineering. We demonstrated that it is possible to distinguish software engineering related search queries using machine learning despite the lack of labeled data. We further performed a thorough analysis on a sample of $1$ million web search sessions comprising of roughly $4$ million search queries to better understand software engineering related search queries and sessions. We showed that software engineering related search queries and sessions constitutes a significant volume, over $2.6\%$, of the overall web search sessions. In addition, we found that software engineering related search tasks are less effective and require more effort than other search queries. Subsequently, we proposed a taxonomy for various user intents, namely - API, Debug, How-to, Learn, Installation, Navigational, and Miscellaneous, for web search in software engineering tasks. Lastly, we conducted an extensive analysis on a sample of $6$ million SE queries to understand the distribution of intents across various web search metrics and other trend characteristics. We believe that these insights will be helpful in improving and maintaining existing tools and building new tools for software engineers.


%
\bibliographystyle{IEEEtran}
\bibliography{references}

\end{document}